\documentclass{article}
\usepackage{amsfonts,amssymb,hyperref}
\usepackage{times,mathptm,anysize}
\usepackage{amsmath,amsthm,graphicx,color}

\begin{document}

\newtheorem{theorem}{Theorem}
\newtheorem{definition}{Definition}
\newtheorem{property}{Property}

\title{Comparison of the Cost Metrics for\\
Reversible and Quantum Logic Synthesis\thanks{This work was
supported in part by PDF and Discovery Grants from the Natural
Sciences and Engineering Research Council of Canada.}}

\author{
      Dmitri Maslov and D. Michael Miller\\
      Department of Computer Science \\
      University of Victoria \\
      Victoria, BC, V8W 3P6, Canada \\
      dmaslov@uvic.ca, mmiller@cs.uvic.ca
}

\newcommand{\gc}[1]{\textcolor{green}{\bf #1}}
\newcommand{\rc}[1]{\textcolor{red}{\bf #1}}
\marginsize{1in}{1in}{1in}{1in}

\maketitle

\begin{abstract}
A breadth-first search method for determining optimal 3-line
circuits composed of quantum NOT, CNOT, controlled-$V$ and
controlled-$V^+$ (NCV) gates is introduced.  Results are presented for simple gate
count and for technology motivated cost metrics.  The optimal NCV
circuits are also compared to NCV circuits derived from optimal
NOT, CNOT and Toffoli (NCT) gate circuits.  The work
presented here provides basic results and motivation for
continued study of the direct synthesis of NCV circuits, and
establishes relations between function realizations in different
circuit cost metrics.

\end{abstract}

\section{Introduction}
Reversible and quantum logic synthesis have attracted recent
attention as a result of advances in quantum and nano
technologies. Many of the proposed synthesis methods, especially
in the area of reversible logic synthesis, assume large gate
libraries where implementation costs of the gates may vary
significantly. However, these methods most often target
minimization of the gate count. Use of such a circuit cost metric
is likely to result in seemingly small circuits, which are in fact
expensive to construct.

The synthesis of circuits composed of NOT, CNOT and
Toffoli (NCT) gates \cite{ar:spmh} and multiple control
Toffoli gates \cite{ws:mp, co:tk, ws:mmd, co:aj, co:k}
has recently been extensively studied.
Converting an NCT or a multiple control Toffoli gate circuit
into one composed of NOT, CNOT, controlled-$V$ and
controlled-$V^+$ (NCV) gates has also been considered \cite{ar:bbcd, co:mymd}.
Further simplification of NCV circuits has also been well
studied \cite{co:mymd}.

It is an important question how close the circuits found by the
above approaches are to optimal.  The direct synthesis of NCV
circuits is also of considerable interest since intuitively one
would expect that to produce better results than the indirect
route via NCT circuits. Finally, we note that observing optimal
circuits for small cases often will shed light on good (if not
optimal) synthesis approaches.

For these reasons, we here present an approach to finding optimal
3-line NCV circuits using various cost metrics.  We compare these
results to those found through mapping optimal NCT circuits.  The
advantage of direct NCV synthesis will be clear even for the
3-line case.  Also, the results clearly demonstrate the difference
between using simple gate count and technology motivated cost
metrics.

The necessary background is reviewed in Section \ref{sec:b}.  A
breadth-first search procedure to find optimal 3-line NCV circuits
is given in Section \ref{sec:sp} and properties identified in those circuits
are discussed in Section \ref{sec:o}.  Section \ref{sec:cc} presents comparative
results to circuits derived from NCT circuits and for various cost
metrics.  The paper concludes with remarks and suggestions for
ongoing research in Section \ref{sec:c}.

\section{Background} \label{sec:b}
We here provide a brief review of the basic concepts required for
this paper. For a more detailed and formal introduction we refer
the reader to \cite{bk:nc}.

A single quantum bit (qubit) has two values, 0 or 1, traditionally
depicted as $|0\rangle$ and $|1\rangle$ respectively.  The state
of a single qubit is a linear combination $\alpha |0\rangle +
\beta |1\rangle$ (also written as a vector $(\alpha, \beta)$) in
the basis $\{|0\rangle,\;|1\rangle\}$, where $\alpha$ and $\beta$
are complex numbers called the amplitudes, and
$|\alpha|^2+|\beta|^2=1$. Real numbers $|\alpha|^2$ and
$|\beta|^2$ represent the probabilities ($p$ and $q$) of reading
the values $|0\rangle$ and $|1\rangle$ upon physical measurement
of the qubit. The state of a quantum system with $n>1$ qubits is
described as an element of the tensor product of the single state
spaces yielding a normalized vector of length $2^n$ called the
state vector. Quantum system evolution results in changes of the
state vector expressible as products of $2^n \times 2^n$ unitary
matrices. This formulation characterizes a transformation but
provides no indication of its implementation cost.

In one common approach, small gates are used as elementary
building blocks with unit cost \cite{bk:nc, ar:bbcd, co:hsyyp,
co:mymd, ar:llklbp}. Among them are NOT ($x \rightarrow \bar{x}$)
and CNOT ($(x,y)\rightarrow (x, x \oplus y)$) gates, the 2-bit
controlled-V gate which changes the target line according to the
transformation given by matrix ${\bf
V}=\frac{i+1}{2}\left(1\;\;-i\atop -i\;\;1\right)$ if, and only
if, the control line is 1. Controlled-$V^{+}$ analogously applies
the transformation ${\bf V^{+}}={\bf V}^{-1}$.  Consult
\cite{bk:nc} for the case where the control line assumes a quantum
(non-Boolean) value. Gates controlled-$V$ and controlled-$V^+$ are
also referred to as controlled-sqrt-of-NOT gates since
$V^2=(V^+)^2=NOT$. In order to implement a Boolean specification,
and assuming an auxiliary line is available in addition to the
minimal set of lines needed for reversibility \cite{ar:mdCAD}, the
set of gates NOT, CNOT, controlled-$V$ and controlled-$V^+$ is
complete. We call this set the {\bf NCV} gates.

\begin{definition}
The {\bf NCV-111 cost} of a circuit composed of NCV gates is the
number of gates in the circuit.
\end{definition}

In an alternative approach \cite{ar:mvbs, co:smb}, it is observed
that any circuit can be composed with single qubit and CNOT gates
and the circuit cost is calculated based on the number of CNOT
gates required. With regards to the NCV library, we next define
NCV-012 cost and motivate our definition by the fact that
controlled-$V$ and controlled-$V^+$ gates require at most 2 CNOT
gates when decomposed into a circuit with single qubit and CNOT
gates (\cite{bk:nc}, page 181).

\begin{definition}
The {\bf NCV-012 cost} of an NCV circuit is linear with
weights 0, 1, 2, and 2 associated with the gates NOT, CNOT,
controlled-$V$ and controlled-$V^+$, respectively.
\end{definition}

The last cost metric that we consider in this paper is motivated
by a recent investigation of the technological (liquid NMR) costs 
of quantum and reversible primitives \cite{ar:llklbp}.

\begin{definition}
The {\bf NCV-155 cost} of an NCV circuit is linear with
weights 1, 5, 5, and 5 associated with the gates NOT, CNOT,
controlled-$V$ and controlled-$V^+$, respectively.
\end{definition}

Finally, we define the multiple control Toffoli gate \cite{ar:tof}.

\begin{definition}
For the set of Boolean variables $\{x_1, x_2, ...,$ $ x_n\}$ the
{\bf Toffoli gate} has the form $TOF(C;T)$, where
$C=\{x_{i_1}, x_{i_2}, ..., x_{i_k}\},\;T=\{x_j\}$ and $C \cap T =
\emptyset$. It maps the Boolean pattern $(x^+_1, x^+_2, ...,
x^+_n)$ to $(x^+_1, x^+_2, ..., x^+_{j-1},$ $ x^+_j\oplus
x^+_{i_1}x^+_{i_2}... x^+_{i_k}, x^+_{j+1}, ..., x^+_n)$.
$C$ will be called the {\bf control} set and
$T$ will be called the {\bf target}.
\end{definition}

The Toffoli gate and its generalizations with more than two
controls form a good basis for synthesis purposes and have been
used by many authors \cite{ws:mp, co:tk, ws:mmd, co:aj, co:k}.
However, Toffoli gates are not simple transformations. Rather they
are composite gates themselves and Toffoli gates with a large set
of controls can be quite expensive \cite{ar:bbcd, co:mymd}. As a
result, the NCV-111 cost of a 100-gate Toffoli circuit with 10
input/output lines can be as low as 100, or may be as high as
15,200. These are, of course, extreme numbers, however, in Section
\ref{sec:cc} we show a better analysis of how the costs may
differ.

\section{The Search Procedure} \label{sec:sp}

Our main tool for the investigation of relations amongst optimal
NCV-111, NCV-012 and NCV-155 implementations and their relation to
the most commonly used reversible circuit cost metric, multiple
control Toffoli gate count, is a search procedure for the optimal
synthesis of NCV circuits for all 40,320 size 3 reversible Boolean
functions. For this problem, we use a prioritized, pruned
breadth-first search. This is because the search tree for simple
full breadth-first search grows too fast for a search to be
accomplished in feasible time and space. This is in contrast to
using a full breadth-first search for NCT optimal synthesis
\cite{ar:spmh}, which is tractable for 3-line circuits with no
search optimizations.

To see the size of the problem, note that the number of base
transformations for NCV gates is 21: 3 transformations involve a
NOT gate, and the use of the gates CNOT, controlled-$V$ and
controlled-$V^+$ each result in 6 transformations. The results in
Table \ref{tab:ncv-012} show that the length of the optimal
implementations can be up to 16. Thus, the number of nodes at one
level of the search tree can be as high as $21^{16}\approx
1.4*10^{21}$.

Breadth first search was previously applied to the synthesis of
optimal circuits for the size 3 reversible functions using NOT,
CNOT, and Toffoli gate library \cite{ar:spmh}. The size of the
bottom level of their search tree (branching factor to the power
of tree depth) is $9^8 \approx 4.3*10^7$ showing that such a
search is significantly simpler than the one we are pursuing and
requires no techniques to reduce the search space or make the
search efficient. In addition, Toffoli gates are not simple
transformations, while synthesis of optimal circuits makes more
sense in terms of simple transformations.  Our search procedure
can find optimal circuits in any weighted gate count metric, not
just in the simple gate count metric used in \cite{ar:spmh}.

An earlier attempt to synthesize optimal NCV circuits \cite{co:yhsp}
was capable of synthesizing optimal implementations of the maximal
cost 7 in one specific cost metric (NCV-011?). The maximal number of synthesized 
functions is 10136 \cite{co:yhsp}, which is about a quarter of the pool size.  
Our program synthesizes optimal implementations for {\em all} 40,320
size 3 reversible functions, and is not tied to a specific costing metric.
Further, our search procedure is {\em significantly} faster: it completes 
the entire search in the lesser time than the search of all optimal 
5-gate implementations in \cite{co:yhsp}.

In our work, we do not allow a quantum gate (controlled-$V$ and
controlled-$V^+$) to have a control line that may at that point in
the circuit take a quantum (non-Boolean) value. We do not have a
mathematical proof that deleting this {\it restriction} will not result
in construction of smaller circuits. However, numerous experiments
indicate that use of a quantum gate with a quantum control line does
not lead to a more efficient circuit than the one we
construct. The above restriction allows us to work with quaternary
logic instead of a continuous values.

The techniques used to reduce the size of the search include:

\begin{itemize}
\item Based on the observation above, we can view a quantum
function as a base reversible Boolean function with a quantum
factor added. In other words, each of the values that may occur in
the truth table is stored as a 2-bit number. Values
0, 1, $V$ and $V^+$ are represented by 00, 10, 01 and 11,
respectively.
\item While our final goal is to synthesize all size
three reversible functions, and none of the quantum (non-Boolean)
functions, the latter still have to be stored and referred to
during the search process. Each such function is stored in a queue
associated with its base reversible Boolean function that can be
uniquely identified from the first digit in the above 2-bit
encoding. The quantum function itself is then identified by a
24-bit number composed of the second bits, that we call the
quantum signature. A quantum function is uniquely defined by the
base Boolean function and its quantum signature.
\item When we
assign a new gate to the existing optimal cascade we never choose
a gate with the same set of controls and targets as the
immediately previous one in the circuit. This is because such a
sequence can always be reduced by template application
\cite{co:mymd}, and thus will not be part of an optimal circuit.
This cuts down the number of gates that one must consider at each
step.
\item We note that in an NCV implementation of a reversible
function one can interchange controlled-$V$ with controlled-$V^+$
gates without changing the function realized, provided all such
gates are interchanged. In our search procedure, this is accounted
for by never using a controlled-$V^+$ gate as the first quantum
gate during construction of an NCV circuit. In this context, the
\emph{first quantum gate} is the one that transforms a Boolean
line to one that can take on quantum values.
\item Once an optimal
implementation of a function is found, we have also found an
optimal implementation for all functions that differ from this one
only by their input-output labeling. Potentially, this accounts
for at most 6 different functions.
\item Once $G_1G_2...G_k$ is a
circuit for a reversible function $f$,
$G_k^{-1}G_{k-1}^{-1}...G_1^{-1}$ is a valid circuit for $f^{-1}$
\cite{co:mymd}. It can be shown that for each metric  considered
in this paper, as well as in any weighted linear type metric
if $G_1G_2...G_k$ is optimal for $f$, then
$G_k^{-1}G_{k-1}^{-1}...G_1^{-1}$ will be an optimal
implementation for $f^{-1}$. From the point of view of the search
for optimal circuits, this means that once an optimal circuit for
$f$ is found, so is an optimal circuit for $f^{-1}$. This
observation would further help to cut down the search space,
however, we have not yet implemented it since our program is fast
enough at present.   It takes approximately 1 minute to synthesize
optimal 3-line NCV circuits in each metric on a single 750 MHz
processor Sun Blade 1000.
\end{itemize}

\begin{figure}[htb]
\begin{center}
\includegraphics[height=20mm]{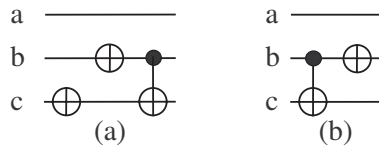}
\caption{(a) Non-optimal, but first found and (b) optimal in NCV-155
cost metric circuits.}
\label{2circ}
\end{center}
\end{figure}

Due to the potentially differing costs of the basis gates, our
procedure maintains several queues of functions, each
corresponding to the cost associated with the circuits in it.
During the search, new gates are assigned to the circuits with
smallest cost not yet considered thereby yielding new circuits to
be considered. Cheaper gates are applied first.  However, due to
varying gate costs, the first circuit found realizing a Boolean
function may not be optimal. To see this, consider the example
illustrated in Figure \ref{2circ}. Our program finds the
non-optimal circuit with NCV-155 cost 7 before the optimal
implementation with NCV-155 cost 6.  This is because the procedure
generates a circuit with two NOT gates before a circuit with a
single CNOT gate, and consequently finds the three gate circuit
in advance of the cheaper two gate alternative. Note that for gate
count these two circuits are generated in the opposite order.

\section{Observations on the optimal 3-line NCV circuits} \label{sec:o}

Using the above search procedure we found several interesting
properties of the optimal 3-line NCV circuits:

\begin{itemize}
\item Optimal implementations found with cost metrics NCV-111 and NCV-155
are interchangeable, and optimal NCV-111 implementations have
optimal costs in NCV-012 cost metric. Diagram in Figure \ref{fig:interchange}
shows which optimal implementations can be substituted without
loosing the property of optimality in the corresponding metric.
Practically, this means that the set of optimal NCV-111 circuits
contains circuits optimal in other (NCV-012 and NCV-155) metrics.
In Section \ref{sec:cc} we make some observations with regard to
the NCV-111 metric optimal circuits. The same comparisons apply
for optimal NCV-012 and optimal NCV-155 implementations. Further,
our experiments with different metrics suggest that the set of
optimal NCV-111 circuits will contain optimal implementations in
NCV-$xyz$ cost metric as long as non-negative integer numbers $x$,
$y$, and $z$ which represent costs of the gates NOT, CNOT, and
controlled-$V$ (and assuming the cost of the controlled-$V+$
equals the cost of the controlled-$V$) satisfy inequality $y<2z$.

\begin{figure}[htb]
\begin{center}
\includegraphics[height=15mm]{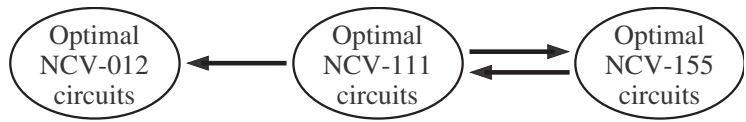}
\caption{Interchangeability of the optimal implementations.}
\label{fig:interchange}
\end{center}
\end{figure}

\item In every circuit cost metric that we have considered (NCV-111, NCV-155,
and NCV-012), the total number of controlled-$V$ and
controlled-$V^+$ gates in any single circuit is divisible by 3 and
is never more than 9. We conjecture that the overall number of
controlled-$V$ and controlled-$V^+$ gates in any NCV
implementation of any reversible function is divisible by three.

\item A 3-bit Toffoli gate and a 3-bit Toffoli gate with one
negative control have the same cost in each of the metrics
NCV-111, NCV-155, and NCV-012. Some optimal circuits are
illustrated in Figure \ref{toffolis}. An optimal implementation
of a 3-bit Toffoli gate with two negative controls has 6 gates.

\begin{figure}[htb]
\begin{center}
\includegraphics[height=20mm]{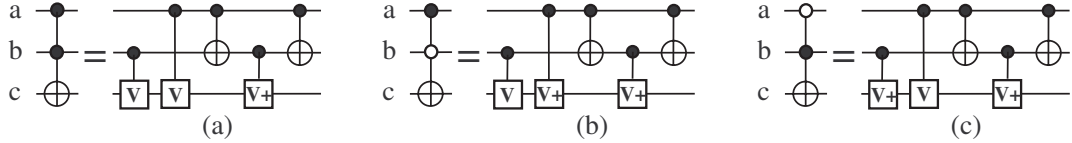}
\caption{(a) optimal NCV realization of the Toffoli gate (b),(c) optimal NCV
realizations of the Toffoli gate with a single negative control.}
\label{toffolis}
\end{center}
\end{figure}

This observation allows us to generalize the famous result by
Barenco {\em et al.} \cite{ar:bbcd} and improve \cite{ar:mdCAD} on
the implementation costs of the generalized multiple controlled
Toffoli gates by suggesting that large Toffoli gates with some,
but not all, negative controls can be simulated with the same cost
as the same size Toffoli gate with all positive controls.
Illustrated in Figure \ref{lt} is a Toffoli gate with 5 controls,
and a circuit simulating it, analogous to the one from
\cite{ar:bbcd}. Since the cost of a size 3 Toffoli gate with one
negative control equals to the cost of size 3 Toffoli gate with
both positive controls, large Toffoli gate illustrated in Figure
\ref{lt} will have a cost equal to the cost of the same size
Toffoli gate but having only positive controls. Application of the
local optimization techniques like \cite{co:mymd} may choose
different scenarios for simplification of the circuits for large
Toffoli gates with all positive and some but not all negative
controls. However, results of such optimization are out of the
scope of the present paper.

\begin{figure}[htb]
\begin{center}
\includegraphics[height=40mm]{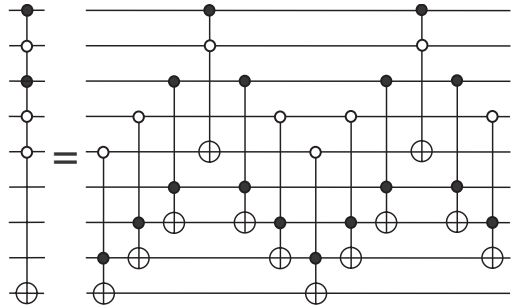}
\caption{Construction of a large Toffoli gate with some but not
all negative controls.} \label{lt}
\end{center}
\end{figure}

\item Observing all optimal implementations in the NCV-111 metric we came to the
conclusion that every Boolean function $f(x_1,x_2)$ of two variables can be
computed by a quantum circuit with no more than 5 gates. If this function
is from Boolean class $0$ (when $f(0,0)=0$), only 4 quantum gates are required.

\end{itemize}

\section{Comparisons of the Sets of Optimal Implementations} \label{sec:cc}
In this section, we compare the NCV-111 (NCV-012 and NCV-155)
costs for the optimal synthesis of size 3 reversible functions
using Toffoli gates up to size 3 (NCT library, \cite{ar:spmh}) with the 
costs of optimal NCV circuits. We note that our implementation of
the breadth first search for the size 3 reversible NCT circuits
may differ from the original as discussed in \cite{ar:spmh}. Due
to the large number of optimal NCT circuits for some functions,
the results shown below may vary slightly depending on the actual
program implementation. Further, the comparison is done through
substitution of every Toffoli gate in an optimal NCT circuit with
a 5-gate NCV circuit (\cite{bk:nc}, page 182), and thus assigning
a cost of 5 in the NCV-111 metric. Gates NOT and CNOT are present
in both libraries, NCT and NCV, and thus require no specific
attention.

\subsection{Optimal NCV-111 vs. Optimal NCT Circuits}

\begin{table}
\begin{center}
\begin{tabular}{|c||c|c||c|}\hline
  & \multicolumn{2}{|c||}{Opt. NCT} & Opt. NCV-111   \\
Cost   & GC \cite{ar:spmh}      & NCV-111  & NCV-111  \\ \hline
0  & 1           & 1        & 1  \\
1  & 12          & 9        & 9  \\
2  & 102         & 51       & 51  \\
3  & 625         & 187      & 187  \\
4  & 2780        & 392      & 417  \\
5  & 8921        & 475      & 714  \\
6  & 17049       & 259      & 1373  \\
7  & 10253       & 335      & 3176  \\
8  & 577         & 1300     & 4470  \\
9  & 0           & 3037     & 4122  \\
10 & 0           & 3394     & 10008  \\
11 & 0           & 793      & 5036  \\
12 & 0           & 929      & 1236  \\
13 & 0           & 4009     & 8340  \\
14 & 0           & 8318     & 1180  \\
15 & 0           & 4385     & 0  \\
16 & 0           & 255      & 0  \\
17 & 0           & 1297     & 0  \\
18 & 0           & 4626     & 0  \\
19 & 0           & 4804     & 0  \\
20 & 0           & 475      & 0  \\
21 & 0           & 106      & 0  \\
22 & 0           & 503      & 0  \\
23 & 0           & 357      & 0  \\
24 & 0           & 4        & 0  \\
27 & 0           & 17       & 0  \\
28 & 0           & 2        & 0  \\ \hline
WA & 5.8655      & 14.0548  & 10.0319  \\ \hline
\end{tabular}
\caption{Optimal NCT and optimal NCV-111 NCV circuits.}
\label{tab:ncv-111}
\end{center}
\end{table}

The leftmost column of Table \ref{tab:ncv-111} refers to the cost
of an implementation of a 3-bit reversible function. The second
and third columns refer to the number of optimal NCT circuits
reported in \cite{ar:spmh}; the fourth column refers to the number
of optimal NCV circuits in the NCV-111 metric found by our search
procedure. The circuit costs used in the last three columns are
NCT gate count, NCV-111 and NCV-111 correspondingly. Table also
\ref{tab:ncv-111} reports the weighted average (WA) for the given
cost metric. The third and fourth columns compare the NCV-111
costs of the optimal NCT and optimal NCV-111 implementations in
cost metric NCV-111 for the size 3 reversible functions. We
observed that the maximal ratio of the cost of one optimal
implementation over the other is $3.375=\frac{27}{8}$. That is,
even for circuits with a small number of inputs/outputs an optimal
Toffoli circuit transformed to NCV can be a factor of 3.375 off
its optimal NCV realization. We also determined that on average,
the optimal NCT circuit is 1.3902 times more expensive (for the
NCV-111 metric) than the corresponding optimal NCV circuit found
by our search procedure. An important question of whether the NCT
and NCV costs are related is addressed by the computing the
correlation between the vectors of NCV-111 costs of the optimal
NCT and optimal (in NCV-111 metric) NCV circuits. The correlation
equals 0.896. We conclude that the costs are reasonably
correlated. Finally, we found it interesting to determine how many
optimal NCT circuits have optimal NCV-111 cost. There are 1,610.

\begin{figure}
\begin{center}
\includegraphics[height=53mm]{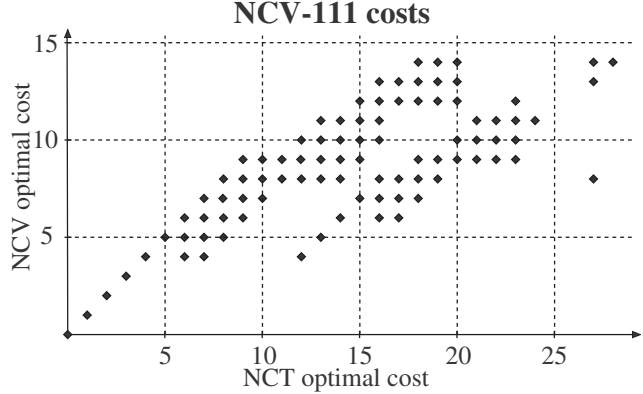}
\caption{Optimal NCT (X-coordinate) VS Optimal NCV-111 (Y-coordinate) circuits.}
\label{ncv-111chart}
\end{center}
\end{figure}

The results are illustrated in the cost comparison chart in Figure
\ref{ncv-111chart}.

\subsection{Optimal NCV-012 vs. Optimal NCT Circuits}

We considered the set of optimal NCV-012 circuits
as created by our program. Optimal NCV-111 implementations were
not substituted for those optimal NCV-012 circuits whose NCV gate
count is not optimal.

\begin{table}
\begin{center}
\begin{tabular}{|c||c|c||c|c|}\hline
  & \multicolumn{2}{|c||}{Opt. NCT} & \multicolumn{2}{|c|}{Opt. NCV-012}  \\
Cost   & GC       & NCV-012  & NCV-012      & NCV-111 \\ \hline
0  & 1           & 8       & 8     & 1 \\
1  & 12          & 48      & 48    & 9 \\
2  & 102         & 183     & 192   & 45 \\
3  & 625         & 398     & 408   & 142 \\
4  & 2780        & 486     & 480   & 315 \\
5  & 8921        & 201     & 192   & 585 \\
6  & 17049       & 16      & 16    & 1169 \\
7  & 10253       & 0       & 192   & 2286 \\
8  & 577         & 47      & 1056  & 3414 \\
9  & 0           & 352     & 3168  & 4790 \\
10 & 0           & 1347    & 4320  & 6744 \\
11 & 0           & 3130    & 672   & 6420 \\
12 & 0           & 3340    & 0     & 4328 \\
13 & 0           & 561     & 0     & 4360 \\
14 & 0           & 3       & 2880  & 4032 \\
15 & 0           & 0       & 11520 & 1568 \\
16 & 0           & 162     & 4416  & 112 \\
17 & 0           & 1219    & 0     & 0 \\
18 & 0           & 4435    & 0     & 0 \\
19 & 0           & 8029    & 0     & 0 \\
20 & 0           & 3872    & 0     & 0 \\
21 & 0           & 128     & 9856  & 0 \\
22 & 0           & 0       & 896   & 0 \\
24 & 0           & 341     & 0     & 0 \\
25 & 0           & 1946    & 0     & 0 \\
26 & 0           & 4482    & 0     & 0 \\
27 & 0           & 3977    & 0     & 0 \\
28 & 0           & 609     & 0     & 0 \\
29 & 0           & 6       & 0     & 0 \\
32 & 0           & 289     & 0     & 0 \\
33 & 0           & 489     & 0     & 0 \\
34 & 0           & 194     & 0     & 0 \\
35 & 0           & 3       & 0     & 0 \\
40 & 0           & 16      & 0     & 0 \\
41 & 0           & 3       & 0     & 0 \\ \hline
WA & 5.8655      & 19.1011 & 14.9800 & 10.5800 \\ \hline
\end{tabular}
\caption{Optimal size 3 reversible circuit NCV-012 costs in NCT and NCV bases.}
\label{tab:ncv-012}
\end{center}
\end{table}

\begin{figure*}
\begin{center}
\includegraphics[height=75mm]{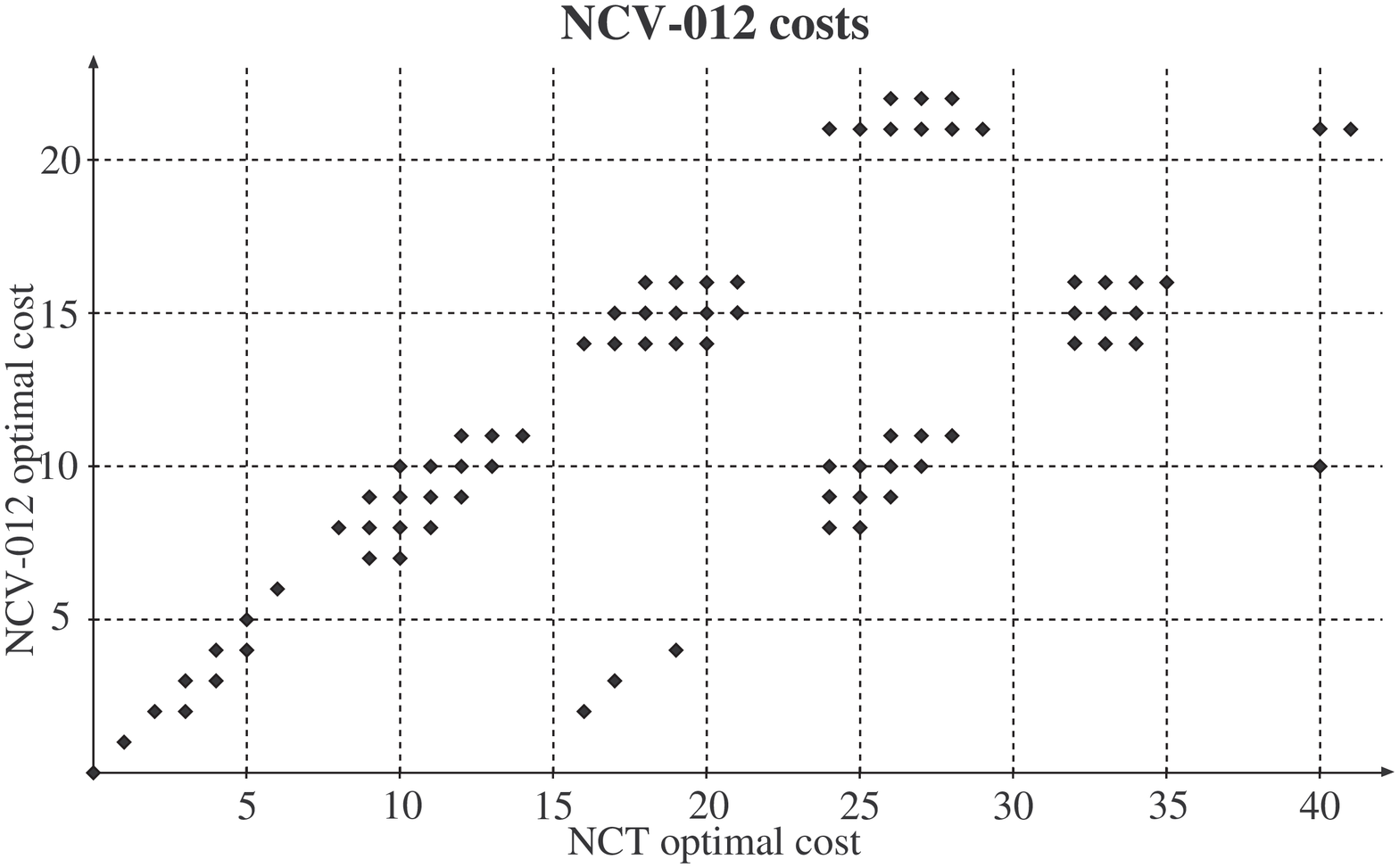}
\caption{Optimal NCT (X-coordinate) VS Optimal NCV-012 (Y-coordinate) circuits.}
\label{ncv-012chart}
\end{center}
\end{figure*}

\begin{figure}[h]
\begin{center}
\includegraphics[height=18mm]{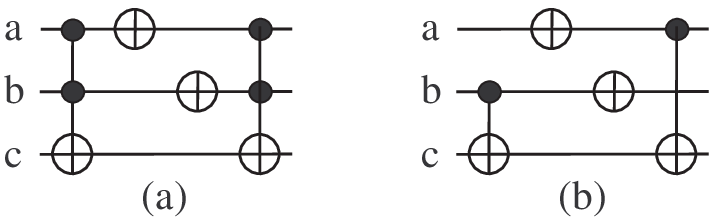}
\caption{(a) optimal NCT and (b) optimal NCV-012 circuits for the function
$[7,6,4,5,2,3,1,0]$.}
\label{max012circ}
\end{center}
\end{figure}

A comparison of the NCV-012 costs of optimal NCV-012 NCV circuits
and NCT circuits is made in Table \ref{tab:ncv-012} (this table is
organized similarly to the previous table) and Figure
\ref{ncv-012chart}. In this metric, the maximum ratio of NCT
optimal circuit NCV-012 cost over NCV-012 optimal circuit NCV-012
cost equals 8 ($=\frac{16}{2}$).  The optimal NCT and optimal
NCV-012 circuits for one such function are illustrated in Figure
\ref{max012circ}). On average, however, this ratio is 1.2728.
Correlation between the vectors of costs is 0.8999, which is
almost identical to the correlation between optimal NCT and
optimal NCV-111 circuits in the NCV-111 cost metric. The number of
functions where the NCV-012 cost of an optimal NCT circuit equals
the NCV-012 cost of an optimal NCV-012 circuit equals 1,774.

%


\section{Optimal implementations with restricted qubit-to-qubit interactions}

In the discussion above we assumed that a direct interaction between any 
two qubit can be established ({\it e.g.} a 2-qubit gate may be built on any 
two qubits). However, due to the specifics of a particular 
physical realization, this may not always be the case. Some of the qubit-to-qubit 
interactions may only be available indirectly. On the other hand,
in every $n$-bit quantum computation it must always be possible to 
construct a connected graph with vertices representing qubits 
and edges representing possibility of the 
direct interaction between qubits.
In the case $n=3$, there are only two non-isomorphic connected graphs.
One is the complete graph, and the remaining is a star (all vertices
are connected to one). 
In this section we report results for the optimal 
synthesis assuming direct interactions are allowed between qubits 
$a$ and $b$, and $b$ and $c$. We assume NCV-111 costing metric, 
however, the results can be calculated for other metrics as well. 
Assuming the implementation costs (numbers of gates) are $0..23$, 
the numbers of functions requiring this many gates are 1, 7, 29, 
82, 181, 334, 374, 334, 337, 753, 1652, 2654, 2482, 1674, 1350, 
3236, 6304, 6028, 1508, 1302, 2566, 4314, 2804, and 14. There are no 
functions requiring more than 23 gates.

\begin{figure}
\begin{center}
\includegraphics[height=27mm]{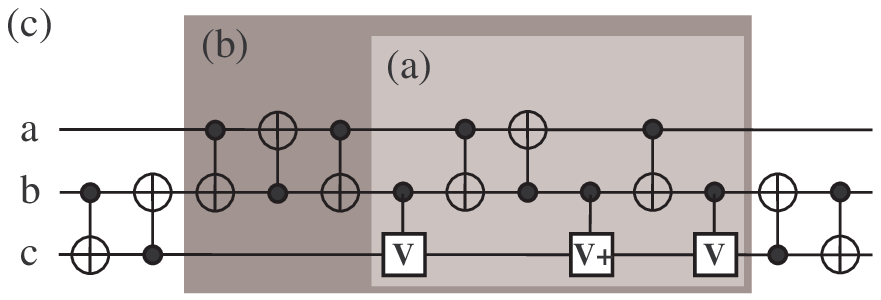}
\caption{Optimal circuits for (a) $(a,b,c)\mapsto(b,a,c\oplus ab)$,
(b) $TOF(a,b;c)$, and (c) $TOF(a,c;b)$.}
\label{cntcirc}
\end{center}
\end{figure}

We found it interesting to notice that in the case of 
non-restricted qubit interactions one of the cheapest non-linear 
(with respect to EXOR) reversible gates 
is the Peres gate \cite{ar:peres} defined by the transformation   
$(a,b,c)\mapsto(a,b\oplus a,c\oplus ab)$. It can be composed with 4 
quantum NCV gates only. An analogy (smallest non-linear reversible gate 
with a similar transformation) of this gate in the case of restricted 
qubit interactions is the gate defined by the transformation 
$(a,b,c)\mapsto(b,a,c\oplus ab)$. It requires 6 quantum NCV 
gates as illustrated in Figure \ref{cntcirc}(a). 
Toffoli gates $TOF(a,b;c)$ and $TOF(b,c;a)$ require 9 elementary quantum 
gates each and can be thought of as a SWAP gate (3 CNOTs) followed by the 
$(a,b,c)\mapsto(b,a,c\oplus ab)$ transformation. An optimal implementation
of $TOF(a,b;c)$ is illustrated in Figure \ref{cntcirc}(b).
Toffoli gate $TOF(a,c;b)$ is somewhat more expensive. It requires 13 
elementary operations in its optimal implementation (see Figure \ref{cntcirc}(c)).

\section{Conclusion} \label{sec:c}

The results in Section \ref{sec:cc} lead us to the following
conclusion. Minimization of Toffoli gate count as a criterion for
a reversible synthesis method is not optimal and even for small
parameters may result in a seemingly small circuit which may be as
far off a technologically favorable implementation as a factor of
8. It is natural to expect that for circuits with more lines the
difference will grow. We suggest that the commonly used gate count
metric should be replaced with a metric that accounts for the
different costs of large building block (i.e., Toffoli gates),
such as the weighted gate cost
presented here. Using such a metric would lead to the
technologically favorable circuit (b) in Figure \ref{max012circ}.

Finally, we observe that for some small parameters minimizing the
cost in any of the technology oriented metrics that we considered
in this paper (NCV-111, NCV-155 or NCV-012) should result in
a reasonably small if not optimal solution if
another metric is applied to the cascade of gates. Thus, at the
present time it is sufficient to use any of the suggested
technology oriented cost metrics.
NCV-111 and NCV-155 are preferred over the NCV-012 metric, since
all such realizations (assuming reversible functions with no more
than 3 variables) have optimal NCV-012 cost.

\end{document}